\documentclass[
12pt,showpacs,preprint,tightenlines,amsmath,amssymb
,superscriptaddress
]{revtex4}
\usepackage{graphicx,epsfig}
\usepackage{dcolumn}
\usepackage{bm}
\usepackage{amsmath}
\usepackage{psfig}
\def\beq{\begin{equation}}
\def\eeq{\end{equation}}
\def\bea{\begin{eqnarray}}
\def\eea{\end{eqnarray}}
\def\beqa{\begin{equation}\begin{array}{l}}
\def\eeqa{\end{array}\end{equation}}
\def\eqlab#1{\label{eq:#1}}
\def\figlab#1{\label{fig:#1}}
\def\tablab#1{\label{tab:#1}}

\def\eref#1{(\ref{eq:#1})}
\def\Eqref#1{Eq.~(\ref{eq:#1})}

\def\Figref#1{Fig.~\ref{fig:#1}}
\def\tabref#1{\ref{tab:#1}}

\def\sla#1{#1 \hspace{-2mm} \slash}

\def\barr{\left(\begin{array}{c}}
\def\earr{\end{array}\right)}
\def\bmat{\left(\begin{array}{cc}}
\def\emat{\end{array}\right)}
\def\al{\alpha}

\def\ga{\gamma} 
 \def\De{\Delta}

 \def\La{{\Lambda}}

\def\w{\omega}

\def\nn{\nonumber}

\def\mathscr{\mathcal}

\def\3d{3-D}

\def\eqlab#1{\label{eq:#1}}
\def\figlab#1{\label{fig:#1}}
\def\tablab#1{\label{tab:#1}}

\def\eref#1{(\ref{eq:#1})}
\def\Eqref#1{Eq.~(\ref{eq:#1})}

\def\Figref#1{Fig.~\ref{fig:#1}}
\def\tabref#1{\ref{tab:#1}}

\begin{document}
\title{$\gamma^{*}N\Delta$  Form Factors from a Relativistic
Dynamical Model of Pion Electroproduction}
\author{G. L. Caia}
\email{caia@phy.ohiou.edu} \affiliation{ Institute of Nuclear and
Particle Physics (INPP), Department of Physics and Astronomy, Ohio
University, Athens, OH 45701}

\author{V.~Pascalutsa}
\email{vlad@jlab.org}
\affiliation{Department of Physics, The College of William \& Mary, Williamsburg, VA 23188}
\affiliation{Theory Group, Jefferson Laboratory, 12000 Jefferson Ave, Newport News,
VA 23606}
\author{J. A. Tjon}
\email{tjon@jlab.org}
\affiliation{Theory Group, Jefferson Laboratory, 12000 Jefferson Ave, Newport News,
VA 23606}
\affiliation{Department of Physics, University of Maryland, College Park, MD 20742}
\author{L. E. Wright}
\email{wright@ohiou.edu} \affiliation{ Institute of Nuclear and
Particle Physics (INPP), Department of Physics and Astronomy, Ohio
University, Athens, OH 45701}

\begin{abstract}
We obtain the electromagnetic form factors of the $\gamma N\Delta$ transition
by analyzing recent pion-electroproduction
data using a fully  relativistic dynamical model.
Special care is taken to satisfy Ward-Takahashi identities for the Born term
in the presence of form factors thereby
 allowing the use of realistic electromagnetic form
 factors of the nucleon and pion. We parametrize the
$Q^2$ dependence of the {\it bare} $\gamma N \Delta$
form factors by a three-parameter form which is consistent with
the asymptotic behavior inferred from  QCD. The parameters of
the bare $\gamma N \Delta$ form factors
 are the only free parameters of the model and are
fitted to the differential cross-section and multipole-analysis
data up to $Q^2=4$ (GeV/c)$^2$ in the $\Delta(1232)$-resonance region.
This analysis emphasizes the significance of the pion-cloud effects
in the extraction of the resonance parameters.
\end{abstract}

\maketitle
Recently a covariant quasipotential description of pion-nucleon ($\pi N$) interaction~\cite{Pascalutsa1}
has been extended to the process of pion photo-production~\cite{Pascalutsa2}.
Here we further extended this description to the case of
pion {\it electro}-production and use the resulting model to
extract the electromagnetic $N$ to $\De$ transition form factors. In doing so
we exploit the results of the recent pion production experiments at LEGS \cite{LEGS}, MAMI
\cite{MAMI},  JLab \cite{Frolov,Joo}, and MIT-Bates \cite{Bates}, as well as  the multipole
analysis MAID~\cite{MAID1}.

One of the aims of our investigation is to disentangle the effect of the
resonance excitation from the  competing background mechanisms, such as pion
rescattering in the final state.
Admittedly, any such separation of resonance and background is model dependent.
We have attempted
to constrain this model-dependence by adhering to general principles, such as
relativistic covariance, current conservation, unitarity, chiral symmetry.
The model is based on a $\pi N$-$\ga N$ coupled-channel
equation which when solved to the first order in the electromagnetic coupling $e$
leads to the electroproduction amplitude, $T_{\pi\ga^\ast}=V_{\pi\ga^\ast}+
T_{\pi\pi} G_\pi V_{\pi\ga^\ast}$, where  $V_{\pi\ga^\ast}$ is
an basic electroproduction potential, $G_\pi$ is the pion-nucleon propagator
and $T_{\pi \pi}$ is the full $\pi N$ amplitude. Thus,  the pion rescattering effects
are included as the final state interaction.
The $\pi N$ amplitude satisfies an integral equation on its own.
The details on reconstructing this amplitude and its fit to the $\pi N$ elastic scattering
are presented in~\cite{Pascalutsa1}.

\begin{figure}[thb]
\centerline{
\includegraphics[totalheight=0.25\textheight]{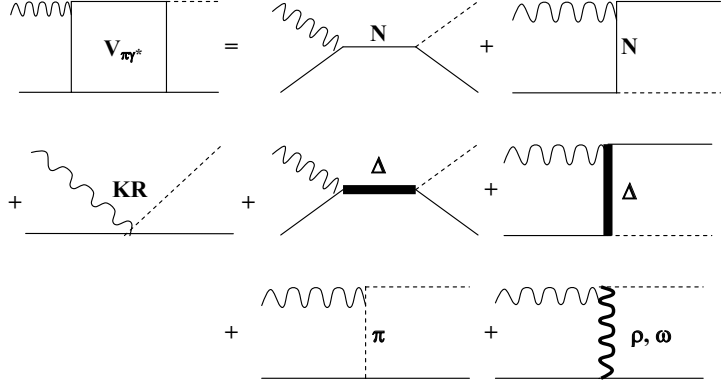} }
\caption{\label{photprod_potential} The Born, vector-meson, and $\De$-isobar
contributions included in the electroproduction potential.} 
\figlab{photprod_potential}
\end{figure}

Our model potential for the pion electro-production is shown in Fig.
\ref{photprod_potential}. It includes the Born term
(we use the pseudo-vector $\pi NN$ coupling  required by chiral symmetry
considerations), the $t$-channel exchange of $\rho-$ and $\w-$ mesons, and
the $\De$-isobar exchange.

The $\pi N$ final state interaction dresses the $s$-channel nucleon
and resonance contributions, leading in particular to the mass, field and
coupling constant renormalizations.
Therefore, both $N$- and $\Delta$-pole contributions in $V_{\pi\ga}$
are included using the {\it bare} mass and coupling
parameters obtained from the equation for the $\pi N$ amplitude.
The renormalization conditions together with unitarity demand that the same
propagators and $\pi N$ vertices, including the cutoff functions,
appear in both the $\pi N$ and $\ga N$ potentials. Thus, all these ingredients
are fixed by the analysis of $\pi N$ scattering~\cite{Pascalutsa1}.

On the other hand, the electromagnetic interaction is constrained by the electromagnetic {\it
gauge invariance}.  At this point one is often concerned with the problem
of how to introduce the electromagnetic form factors for nucleon and pion in a way consistent
with gauge invariance.
A common, but not viable, solution to this problem, implemented for instance
in Refs.~\cite{MAID1,Sato2,DMT},
is to choose all of the electromagnetic form factors
that go into the Born term
 (i.e., nucleon, pion and axial form factors) to be {\it the same}. This  prescription
does enforce the current conservation,  however the Ward-Takahashi (WT)
identities cannot be satisfied in this way. Furthermore, it is clear that
the requirement of gauge invariance should not be able to  restrict the $Q^2$ behavior
of the electromagnetic interaction (see, {\em e.g.,} Ref.~\cite{KPS02}). Finally,
it is well known, especially in view of the new JLab experiments~\cite{ele_proton_ff},
that these form factors {\it are not} the same.

For these reasons we sought for a solution that permits arbitrary choices of
the electromagnetic form factors, yet is fully consistent with gauge invariance.
Following the arguments given in, e.g., \cite{KPS02,GrR87},
we find that an arbitrary form factor $F(Q^2)$ can be accommodated by the
following replacement the current:
\beq
\eqlab{procedure}
J^\mu \rightarrow {J'}^\mu(Q^2)= J^\mu +[F(Q^2)-1] \,O^{\mu\nu} J_\nu\,,
\eeq
where $O ^{\mu \nu}=g^{\mu \nu}-q^{\mu} q^{\nu}/q^{2}$, and $q$
is the photon 4-momentum, $Q^2=-q^2$.
It is easy to see that the resulting
current ${J'}^\mu$ obeys exactly the same WT identities as $J^\mu$.
Thus, as long as gauge-invariance is implemented
at the real-photon point, the inclusion of the form factors via \Eqref{procedure}
will give the gauge-invariant current for $Q^2\neq 0$.
For example, the bare $NN\ga$ and $\pi\pi\ga$ vertex functions and the Kroll-Rudermann term are:
\begin{subequations}
\eqlab{modEMvert}
\begin{eqnarray}
\eqlab{modgaNNvert} \Gamma_{
NN\ga}^{\mu}&=&e\,\gamma^{\mu}+e\,[F_{1}(Q^2)-1] \,O^{\mu \nu}
\gamma_{\nu}+\frac{e\kappa_N}{2m_{N}}\, F_{2}(Q^{2})\,i\sigma ^{\mu \nu}q_{\nu} \,,\\
\eqlab{modgapipivert} \Gamma_{\pi\pi\ga}^{\mu}
&=&e\,(k+k')^{\mu}+e\,[F_{\pi}(Q^2)-1] \,O^{\mu \nu}(k+k')_{\nu}\,,\\
\eqlab{KRmodchannel} J_{KR}^{\mu}&=&\frac{e g_{\pi
N}}{2m_{N}}\left\{
\gamma^{\mu}+[F_{A}(Q^{2})-1]\,O^{\mu \nu}\gamma_{\nu}\right\}\gamma_{5}\,.
\end{eqnarray}
\end{subequations}

The above procedure allows us to use the experimentally
determined form factors in the Born terms.  Since many of the form factors are of the dipole form we
introduce:
$ F_{D}(Q^{2}; \Lambda^2)=(1+Q^{2}/\Lambda^2)^{-2} $.  The
newly measured {\it proton electric} form factor \cite{ele_proton_ff},  we
represent by the following form:
$G_{E}^{p}(Q^{2})=(1+Q^{2}e^{-Q^{2}})\,F_{D}(Q^{2};0.4)$; it is in a very good agreement
with the more common form~\cite{Vdh02} in the range of $Q^2$ under consideration.
 For the
{\it neutron electric} form factor we use the
parametrization of Galster~\cite{ele_neutron_ff}. For the {\it magnetic} form
factors of both {\it proton and neutron} we use the dipole
form: $G_{E}^{p/n}(Q^{2})=\mu_{p/n}F_{D}(Q^{2},0.71)$.
For the {\it pion} form factor we use the monopole form:
$F_{\pi}(Q^{2})=(1+Q^{2}/0.45)^{-1}$, while, for the axial
form factor we use: $F_{A}(Q^{2})=F_{D}(Q^{2},0.9)$. For
the vector mesons ($\rho/\omega$) we use the prediction of \cite{rho_om_ff}:
$F_{\rho/\omega}=(1+Q^{2})/(1+3.04Q^{2}+2.42Q^{4}+0.36Q^{6})$.

The only undetermined form factors in our model
are then  the $\ga N\Delta$ form factors.  These we 
determine within the framework of our model by fitting to the MAID multipole analysis of  data
in the region of the $\De$-resonance.

Let us first consider the $\ga N\De$ vertex function.
We write it in the Lorenz-covariant form that obeys both electromagnetic
and spin-3/2 gauge symmetries (see~\cite{Pascalutsa3,Pascalutsa4,Pascalutsa5} for details):
\begin{widetext}
\begin{eqnarray}
\eqlab{newvf}
\eqlab{gaNDVlad}  \Gamma_{\gamma \Delta N}^{\alpha
\mu}(p,q) &=&
-\frac{3e\,(m_{\Delta}+m_{N})}{2m_{N}[(m_{\Delta}+m_{N})^{2}-q^{2}]}
\left\{ g_M(Q^{2})\,
\varepsilon^{\alpha \mu \beta \nu} p_{\beta} q_{\nu} \right. \nonumber \\
&+& g_E(Q^{2}) \,(p\cdot q g^{\alpha
\mu}-q^{\alpha}p^{\mu})\,i\gamma^{5} \\
&+&\left. g_C(Q^{2})\,(1/m_{N})\,[q^{2}(p^{\mu}
\gamma^{\alpha}-\sla p g^{\alpha \mu})+q^{\mu}(\sla p
q^{\alpha}-p\cdot q \gamma^{\alpha})]\,i\gamma^{5}\right\},\nn
\end{eqnarray}
\end{widetext}
where $q$ ($\mu$) and $p$ ($\al$) are the four-momenta  (vector-indexes) of the photon and
$\Delta$ respectively.
This  vertex function has the advantage of decoupling the unphysical spin-$1/2$ sector
of the spin-3/2 field, which can be viewed as the result of
the transversality property~\cite{Pascalutsa4}:
$p_{\al} \Gamma_{\gamma \Delta N}^{\alpha\mu}=0$.

At the $\De$-pole (the mass shell of the $\De$),
we can relate these couplings to the more conventional decomposition
of Jones and Scadron~\cite{Scadron} which is done
in terms of $G_M$,$G_E$, and $G_C$ where $M, E,$ and $C$ refer to magnetic,
electric and Coulomb $\ga N\De$ form factors, similar to  {\it Sachs} form factors of the nucleon.
Defining, $D(Q^2)=Q^2+(m_{\Delta}-m_N)^2$
and $P(Q^2)=m_{\Delta}^2-m_N^2-Q^2$, we find the following
 relation between the two sets of form factors:
\begin{subequations}
\eqlab{ff_Vlad}
\begin{eqnarray}
\eqlab{G1_Vlad} g_M&=&G_{M}-G_{E}\\
\eqlab{G2_Vlad}
g_E&=&\frac {2}{D} [P\, G_E + Q^2 \,G_C]\\
\eqlab{G3_Vlad}
g_C&=&\frac{m_N}{m_{\Delta} D}[4 m_{\Delta}^2 \,G_E-P\,G_C]
\end{eqnarray}
\end{subequations}

In the actual calculations we use the vertex function \eref{newvf} with form factors
expressed in terms of $G_{M}$, $G_{E}$ and $G_{C}$ via \Eqref{ff_Vlad}. The $\Delta$ contribution
to the resonant multipoles $M_{1+}^{3/2}$, $E_{1+}^{3/2}$ and $S_{1+}^{3/2}$ are directly proportional to
 $G_{M}$, $G_{E}$ and $G_{C}$, respectively.

To parametrize the $\ga N\De$ form factors we universally use the form:
\beq
\eqlab{gg}
G_I (Q^2) = G_{I}\,\frac{1+(Q^{2}/A_I)
\, e^{-Q^2/B_I} }{(1+Q^{2}/\Lambda_I^2)^2}\,,\,\,\,\,\,I=M,\,E,\,C.
\eeq
Here we have built in a constraint from  perturbative QCD (pQCD) such that
these form factors fall as $Q^{-4}$ (modulo logs) for asymptotically
large $Q$, see e.g.~\cite{Carlson98}. This is the pricipal  difference with parametrizations of 
Refs.~\cite{Sato2,DMT} which fall off exponentially and hence do not satisfy the pQCD constraint.

\begin{figure}[htb]
\begin{center}
\includegraphics[totalheight=0.27\textheight]{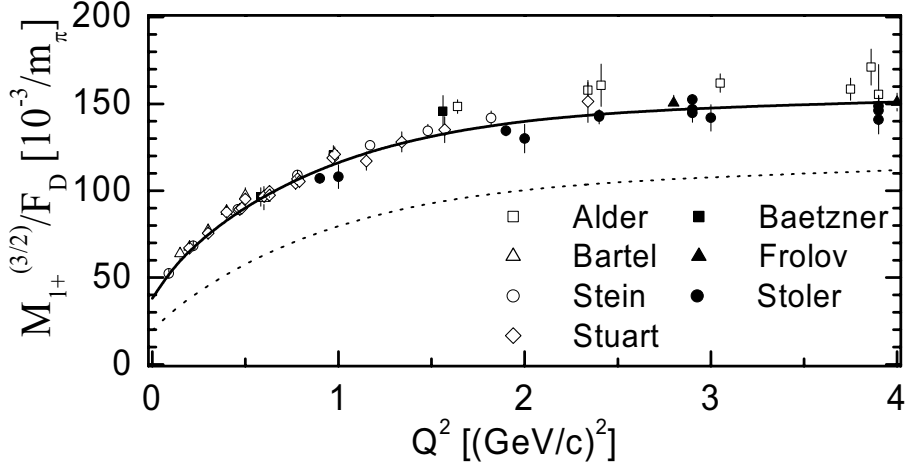}
\caption{\label{M1} Model description of Im$M_{1+}^{3/2}/F_{D}$, with
$F_{D}=(1+Q^{2}/0.71)^{-2}$. The
data at $Q^{2}=2.8$ and $4.0$ $(GeV/c)^{2}$ are taken from \cite{Frolov},
other data are from \cite{M1data}. The dotted line is the calculation including only the
direct $\De$ exchange in $V_{\pi\ga}$, hence leaving out resonant mechanisms due to 
the $\pi$N rescattering.}
\end{center}
\figlab{M1+}
\end{figure}

The photo-couplings, i.e., $G_{M}$ and $G_E$, are
determined by the photoproduction multipoles $M_{1+}^{3/2}$ and $E_{1+}^{3/2}$
at the resonance position $W\simeq 1232$ MeV.
The strength of $G_C$ is determined by $S_{1+}^{3/2}$
at low $Q^2$.
We then have determined the $Q^2$ dependence
of $G_M$ by  comparing our calculations to the experimentally extracted $M_{1+}$ multipole at
an invariant energy $W=1232$ MeV, see Fig.~\ref{M1}. The solid line here represents
our full model calculation of this multipole.
The dotted line represents the result obtained 
by including only the $\Delta$ $s$-channel exchange in the electroproduction potential.
In doing so we exclude mechanisms of resonance electroexcitation through the pion cloud,
see, e.g., \Figref{mechms}. In our model these mechanisms result from the $\pi N$ rescattering through
the resonant channel.  Such mechanisms apparently account for about 50\% of the
$M_{1+}$ strength near the photon point and about 25\% near $Q^2=4$ GeV$^2$. This is
in a qualitative agreement with the findings of Refs.~\cite{Sato2,DMT}.

The parameters of the electric and Coulomb form factors
are adjusted for the best description of the corresponding resonant multipoles known from MAID
analysis. The values of the extracted $\ga N\De$ parameters are summarized
in Table~\tabref{params}.

The ratios of the resonant  multipoles: $R_{EM}=\mbox{Im}E_{1+}/\mbox{Im} M_{1+}$ and
$R_{SM}=\mbox{Im} S_{1+}/\mbox{Im} M_{1+}$ carry important information
about the $N$ to $\Delta$ transition and about the admixture
the $D$-wave component in the nucleon wave function in particular.
The focus of several recent theoretical~\cite{Sato2, DMT,MAID2,Azn02} and experimental
\cite{Frolov,Joo} studies has been the extraction of the $Q^2$-dependence of
these ratios. This dependence can potentially tell us about the range of the momentum transfer where
pQCD becomes applicable (pQCD predicts $R_{EM} = 100 \%$ and $R_{SM} = const$).

\begin{figure}[htb]
\figlab{REM_RSM}
\begin{center}
\includegraphics[totalheight=0.3\textheight]{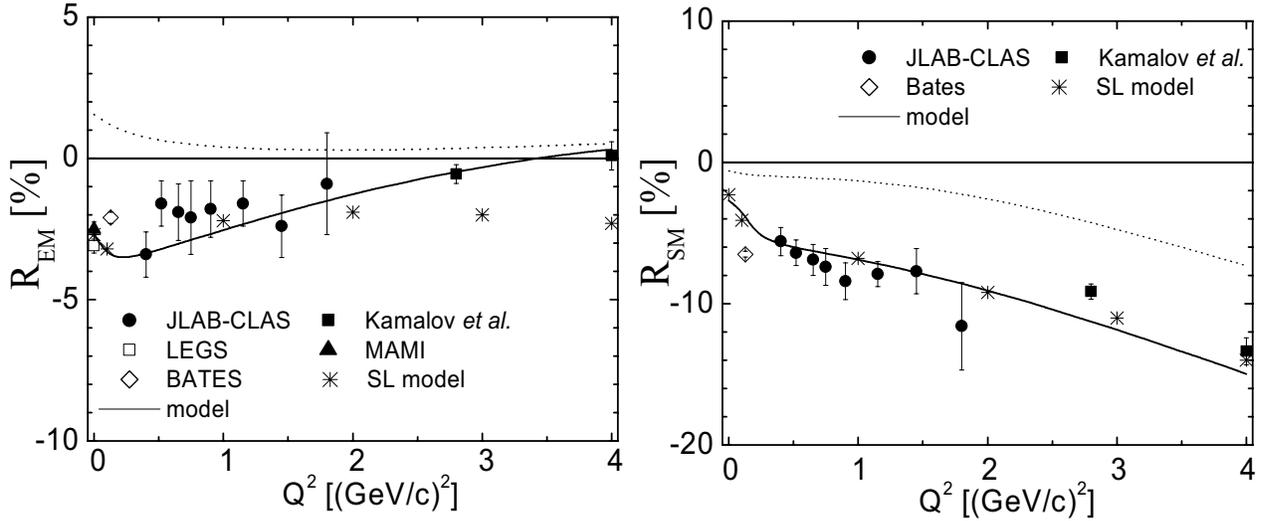}
\caption{\label{EM_SM} Results for the ratios of the resonant multipoles
$R_{EM}$ and $R_{SM}$.  The data points are from CLAS~\cite{Frolov,Joo},
LEGS~\cite{LEGS}, MAMI~\cite{MAMI}, MIT-Bates~\cite{Bates}, and
the MAID reanalysis of JLab data by Kamalov {\it et al.}~\cite{MAID2}. The asterisks represent
predictions of the Sato and Lee model~\cite{Sato2}.}
\end{center}
\end{figure}

In Fig. \ref{EM_SM} we display
the $Q^2$-dependence of $R_{EM}$ and $R_{SM}$ obtained in our model and compared
with a number of results of recent measurements and calculations.
In particular, at the real photon point we find
$R_{EM}\cong -2.7 \%$ and
$R_{SM}\cong-2.3 \%$.
As for the $Q^2$ dependence, we can see that
 $R_{EM}$ shows a systematic tendency
to cross zero in the region between 3 and 4 (GeV/c)$^2$ . This is in
contrast to the recent data analysis \cite{Frolov} or
Sato and Lee model~\cite{Sato2} which conclude that $R_{EM}$ stays
negative and is virtually flat in this domain of
$Q^{2}$.  Our results are  in a better agreement with the
data analysis of Kamalov {\it et al.} \cite{MAID2} and Aznauryan~\cite{Azn02}, which
indicate the sign change of $R_{EM}$ below $Q^{2}=4$
(GeV/c)$^{2}$.   Additional experiments and extractions of
multipoles at higher $Q^2$ would be desirable to investigate this point.
Our result for $R_{SM}$ is in agreement with most of the mentioned analyses.
Again the dotted curves represent the result of resonance dominance, when
$R_{EM}=-G_E/G_M$ and $R_{SM}=G_C/G_M$. 

\begin{figure}[ht]
\begin{center}
\includegraphics[totalheight=0.7\textheight]{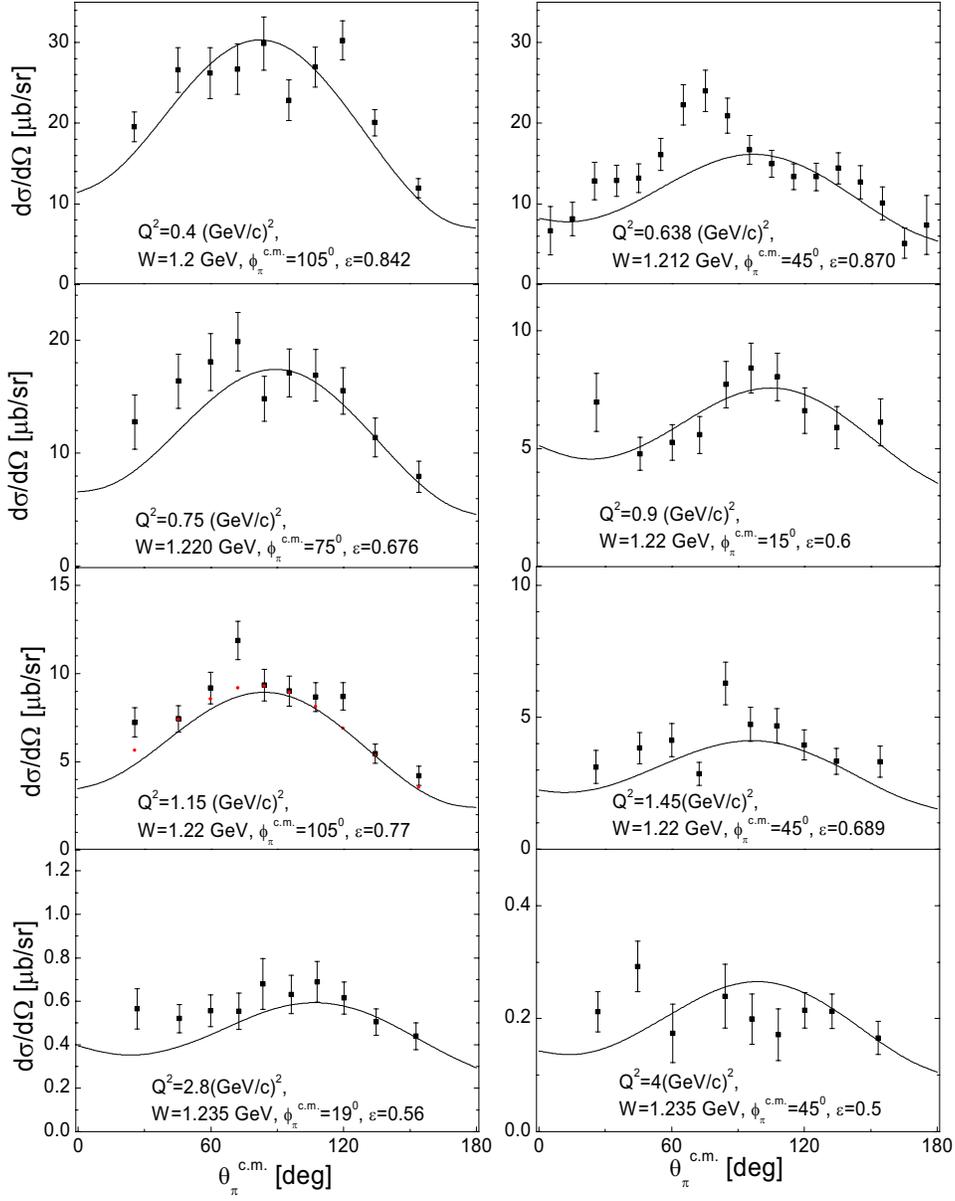}
\caption{\label{dsig_pi0p} Results for the virtual-photon differential cross
sections on $p(e,e'p) \pi^{0}$. The data points are from Refs.~\cite{Frolov,Joo}.}
\end{center}
\figlab{dsig_pi0p}
\end{figure}

After determining the $Q^2$
dependence of the $\ga N\Delta$ form factors, we
calculated the virtual-photon differential cross sections,
spanning a wide range of $Q^{2}$ values as well as various energies $W$.
In Fig.~\ref{dsig_pi0p} we show several of these
calculations. We find the agreement with the recent JLab experiments quite pleasing,
especially in view of the fact that the model has very few adjustable parameters, all given in
Table~\tabref{params}.

In summary, we have extended  the relativistic dynamical model of
Refs.~\cite{Pascalutsa1, Pascalutsa2} to calculate the pion
electro-production reactions. We were able to use realistic electromagnetic
form factors at each photon vertex by carefully treating the
problem of gauge invariance for the Born terms. For the resonance
terms  we parameterized the {\it bare} $\ga N\Delta$
form factors such that the pQCD asymptotic $Q^2$ constraint is fulfilled, and
then fitted them to the cross-section data and resonant multipoles of MAID
up to 4 (GeV/c)$^2$.   Our result for $R_{EM}$ in this  $Q^{2}$ domain is
still very far from the pQCD prediction of $100 \%$.
Although,  we do find that $R_{EM}$
shows a strong tendency to cross zero and change sign in the region between
$3$ and $4$  (GeV/c)$^{2}$. Our model calculation of the virtual photon cross section of
$p(e,e'p) \pi^{0}$ shows a very good
agreement with recent data up to $4$ (GeV/c)$^2$ in the $\De$ resonance energy region.
A thorough study of the existing data base of pion electroproduction reactions using
this model is underway~\cite{Caia04}.

We thank Marc Vanderhaeghen for helpful discussions.
This  work is supported by the DOE under the contracts DE-FG02-93ER40756,
DE-FG05-88ER40435, DE-FG02-93ER-40762, DE-AC05-84ER-40150,
  and the NSF under grant
NSF-SGER-0094668.

\newpage 

\begin{table}[htb]
\begin{tabular}{||l|r|r|r||}
\hline\hline
 & \quad $G_M(Q^2)$ & \quad $G_E(Q^2)$ & \quad $G_C(Q^2)$  \\
\hline
$G(0)$ & 3.10 & 0.05 &$-0.18$\\
\hline
$\La^2$ [GeV$^2$] & 0.6 & 0.5 & 0.8\\
\hline
$A$ [GeV$^2$] & 1.0  & $-1.1$ & $-0.9$\\
\hline
$B$ [GeV$^2$] & 1.2 & 2.0 & 1.0\\
\hline\hline
\end{tabular}
\caption{Parameters of the $\ga N\De$ form factors extracted in our model.}
\tablab{params}
\end{table}



\begin{figure}[thb]
\centerline{
\includegraphics[totalheight=0.05\textheight]{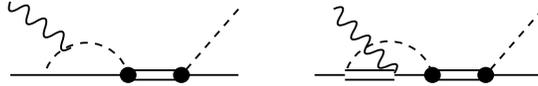} }
\caption{ Examples of the dynamical mechanisms of $\De$-resonance electroexcitation.} 
\figlab{mechms}
\end{figure}

\end{document}